\documentclass[final,sort&compress,5p,times,twocolumn]{elsarticle}

\usepackage{amssymb}
\usepackage{caption}
\usepackage{array}
\usepackage{rotating}
\usepackage{subfigure}
\usepackage{multirow}
\usepackage{color}
\usepackage{epsfig}
\usepackage{epstopdf}
\usepackage{url}
\usepackage[linesnumbered, ruled, vlined]{algorithm2e}
\usepackage{pifont}% http://ctan.org/pkg/pifont
\newcommand{\xmark}{\ding{55}}%

\journal{Computers \& Security}

\begin{document}

\begin{frontmatter}



\title{Multivariate Big Data Analysis for Intrusion Detection: 5 steps from the haystack to the needle}


\author{Jos\'e Camacho}
\author{Jose Manuel Garc\'ia-Gim\'enez}
\author{Noem\'i Marta Fuentes-Garc\'ia}
\author{Gabriel Maci\'a-Fern\'andez}

\address{Department of Signal Theory, Telematics and Communications \\
		School of Computer Science and Telecommunications - CITIC\\
		University of Granada (Spain)\\}
		
\cortext[cor1]{Corresponding author: J. Camacho (email: josecamacho@ugr.es)}

\begin{abstract}
	The research literature on cybersecurity incident detection \& response is very rich in automatic  detection methodologies, in particular those based on the anomaly detection paradigm. However, very little attention has been devoted to the diagnosis ability of the methods, aimed to provide useful information on the causes of a given detected anomaly. This information is of utmost importance for the security team to reduce the time from detection to response. In this paper, we present Multivariate Big Data Analysis (MBDA), a complete intrusion detection approach based on 5 steps to effectively handle massive amounts of disparate data sources. The approach has been designed to deal with the main characteristics of Big Data, that is, the high volume, velocity and variety. The core of the approach is the Multivariate Statistical Network Monitoring (MSNM) technique proposed in a recent paper. Unlike in state of the art machine learning methodologies applied to the intrusion detection problem, when an anomaly is identified in MBDA the output of the system includes the detail of the logs of raw information associated to this anomaly, so that the security team can use this information to elucidate its root causes. MBDA is based in two open software packages available in Github: the MEDA Toolbox and the FCParser. We illustrate our approach with two case studies. The first one demonstrates the application of MBDA to { semistructured} sources of information, using the data from the VAST 2012 mini challenge 2. This complete case study is supplied in a virtual machine available for download. In the second case study we show the Big Data capabilities of the approach in data collected from a real network with labeled attacks. 
\end{abstract}


\end{frontmatter}



\section{Introduction}
\label{sec:introduction}

Intrusion Detection has become a priority for many organizations, as a result of data becoming a fundamental asset for creating value on modern services. According to the VERIZON annual ‘Data Breach Investigation Report’ (DBIR) \cite{DBIR}, in 2017 several tens of thousands of attacks targeted private and public corporations. The 94$\%$ of these attacks had an economic motivation, including industrial espionage. They show a profile of high sophistication, with organized teams of high technical specialization that are sometimes referred to as Advanced Persistent Threats (APTs). An alarming factor of the DBIR report is the ratio between mean time of compromise and mean time of detection. While the former is in the order of minutes, the {latter} is in the order of weeks or even months. This means the attacker has plenty of time for privilege escalation and any other necessary malicious activity to access and ex-filtrate protected data. On average, stolen personal registers are in the tens of millions per attack, with a tremendous impact on the corporate image. To mention some notorious examples, the data breach of  412M personal files including email accounts associated to the adult dating service Friend Finder Network and Penthouse.com and the data breach of more than a billion personal accounts from the email operator River City Media, used in a massive SPAM campaign \cite{IIB}.       

This situation has boosted the market of intrusion detection, in particular of the so-called Security Information and Event Management (SIEM) systems \cite{SIEM}. A SIEM is aimed at aggregating and analyzing data coming from diverse sensor devices deployed through the network, with the ultimate goal of detecting, triaging and validating security incidents. In a recent report \cite{SIEMMarket}, Gartner estimates in 90 billion dollars the  cost associated to cybersecurity incidents in 2017, which implies an increment of $7.6\%$ with respect to the preceding year. The report foresees a change of tendency in cybersec, with a growing effort in detection and response in comparison to prevention means. Following this tendency, the corporations are devoting more economic and human resources to incident detection, creating the so-called Cyber Incident Response Teams (CIRTs). A CIRT devotes a large proportion of its resources to analyze potential incidents. A main limitation for that is the shortage of specialized professionals. Combining this shortage with the proliferation of information leakage attacks, there is a clear need of efficient tools and mechanisms to aid in the detection, triaging and analysis of incidents, in order to make CIRTs more effective in the arms race against APTs.

Multivariate Analysis has been recognized as an outstanding approach for anomaly detection in several domains, including industrial monitoring \cite{Ferrer2014} and networking \cite{Fernandes2016}. In the field of industrial processing, in particular in the chemical and biotechnological industries, the state-of-the-art multivariate approach for anomaly detection is named Multivariate Statistical Process Control (MSPC), and has been developed for more than three decades. In a previous work \cite{MSNM2015}, we introduced a methodology named Multivariate Statistical Network Monitoring (MSNM), which is an extension of MSPC to the cybersecurity domain. We also extended the multivariate methodology to combine traffic data with sources of security data, like IDS or firewall logs, so as to efficiently integrate disparate sources in the incident detection. 

This paper presents the Multivariate Big Data Analysis (MBDA), a complete intrusion detection and analysis approach based on 5 steps to effectively handle tons of disparate data sources in cybersecurity. The core of the approach is the MSNM technique. MBDA is a Big Data extension of MSNM in which, when an anomaly is identified, the output includes the logs of raw information associated with it. These, in turn, can be presented to the  CIRT, so as to elucidate the root causes for the anomaly. This diagnosis ability of MBDA is a main advantage over other machine learning methodologies.

MBDA is based in two open software packages available in Github: the MEDA Toolbox \cite{Camacho201549} (https://github.com/josecamachop/MEDA-Toolbox) and the FCParser (https://github.com/josecamachop/FCParser), the latter presented in this paper for the first time. The FCParser is a python tool for the parsing of both structured and unstructured logs. With the MEDA Toolbox, multivariate modeling and data visualization of the Big Data stream are possible. Combining these two software packages in the 5 steps of MBDA, we show that we are capable of accurately identifying the original raw information related to a detected anomaly in the Big Data. That is, we find the needle in the haystack. This makes MBDA a perfect tool to dive into the overwhelming volumes of disparate sources of information in the cybersecurity context.     

The rest of the paper is organized as follows. Section \ref{sec:related} provides a brief revision of principal works on PCA-based anomaly detection in networking and extensions to Big Data.
Section \ref{sec:statistical} presents the fundamentals of the approach of this paper.  
Sections \ref{sec:case2}  and \ref{sec:case} illustrate the MBDA approach in two case studies: the VAST 2012 mini challenge 2 and data collected from a real network with labeled attacks. Section \ref{sec:conclusion} summarizes the main conclusions of the work.

% RELATED WORK
\section{Related Work}
\label{sec:related}

Among several other anomaly detection paradigms, statistical solutions have been widely adopted in the literature~\cite{Om12}.
In particular, the use of multivariate approaches such as PCA were proposed more than a decade ago \cite{Kanaoka03}. %As other one-class classifiers \cite{Heller2003}, the main advantage of PCA is its unsupervised nature, which does not require --and is not limited by-- an a-priori specification of potential anomalies in the system. This means that PCA is still useful to detect new types of anomalies, something mandatory in real world network anomaly detection.
{One of the main advantages of PCA, similar to} other one-class classifiers {like OCSVM} \cite{Heller2003}, is its unsupervised nature, which does not require --and is not limited by-- an a-priori specification of potential anomalies in the system. This means that PCA is useful to detect {both known and} new types of anomalies, something mandatory in real world network anomaly detection.

The most referred work for PCA anomaly detection is that of Lakhina et \textit{al.} \cite{Lakhina2004}.
However, a number of flaws are pointed out for this approach \cite{Ringberg2007}, mainly as a result of its differences with the  more developed MSPC theory \cite{MSNM2015}. Several modifications of the {original} approach have been proposed  with the aim of solving some of those {issues}%flaws
, \textit{e.g.} \cite{Callegari2011,Delimargas2014,Callegari2014}. More recent research on multivariate analysis for security-related anomaly detection has opted for combining PCA with other detection schemes. Thus, Aiello et \textit{al.} \cite{Aiello2016} combine PCA with mutual information for profiling DNS tunneling attacks. Fernandes et \textit{al.} \cite{Fernandes2016} combine PCA with a modified version of dynamic time warping for network anomaly detection. They also propose an alternative approach based on ant colony optimization. Jiang et \textit{al.} \cite{Jiang2015} apply PCA over a wavelet transform of the network traffic for network-wide anomaly detection. Chen et \textit{al.} \cite{Chen2016} use a similar approach with multi-scale PCA. {Xia et \textit{al.} propose an algorithm based in the Singular Value Decomposition (SVD) which is combined with other techniques for anomaly detection by considering the cyclostationarity of the data \cite{Xia2018}}. 

Despite such a big effort in the field, most of the proposals still share {part} %some
of the problems {reported related to} %of
\cite{Lakhina2004}. %For this reason, we introduced the MSNM methodology in \cite{MSNM2015} to deal with the network anomaly detection problem by mimicking MSPC. This methodology allows to combine traffic data with other {security data} sources \cite{Camacho2014}, demonstrating  detection capabilities comparable to state of the art machine learning methodologies but with the advantage of providing some diagnosis support \cite{432017}.
{This motivated us to develop} the MSNM methodology{, which was introduced  in 2015 to solve these open problems} \cite{MSNM2015}. {The MSNM methodology} allows to combine traffic data with other {security data} sources \cite{Camacho2014}, demonstrating detection capabilities comparable to {state-of-the-art} machine learning methodologies {with the additional} advantage of providing diagnosis support \cite{432017}.

On the other hand, the huge amount of machine-generated data in the network has fostered the application of Big Data techniques for network anomaly detection \cite{Mikel2017Auth}. A general trend is to apply traditional anomaly detection on top of Big Data tools like Hadoop \cite{Bialecki2005}\cite{Dean2008}, Apache Spark \cite{Zaharia2010}\cite{Zaharia2016} or tools derived from them \cite{Fontugne2014}. Most works are based on clustering and classification techniques \cite{Dromard2005}\cite{Gupta2016}\cite{Gonsalves2015}\cite{Hurst2014}\cite{Rathore2016}\cite{Wallace2016} or use host mining \cite{Xu2009}\cite{Gonsalves2015}\cite{Hadziosmanovic2012} to distinguish between normal and anomalous patterns. A combination of clustering with k-means and host mining is proposed in \cite{Yen2013}. The k-means algorithm is also used in other approaches, such as \cite{Therdphapiyanak2013a}\cite{Therdphapiyanak2013b}. Entropy calculation to identify disturbances in the data is used in \cite{Tian2015}\cite{Wang2016}\cite{Krotofil}. Finally, PCA is used in \cite{Xu2009} for host mining together with classification trees. \\
Most of previous solutions are suited to handle the high volume in Big Data. However, other well-known Big Data problems are the velocity, the veracity and the variety, that is, the high incoming pace of data, the level of trust in data and the requirement to combine disparate data sources, respectively. The velocity is a relevant problem because most of the approaches need to re-build the models on-line from new incoming data. {This problem is dealt in \cite{HONG2015} by proposing an Iterative-Tunning Support Vector Machine (SVM), which makes the training of the model faster than the regular SVM.} %Potential solutions to this problem are \cite{Wallace2016} and \cite{Wang2016}. The veracity is also faced up in \cite{Krotofil2016}. In \cite{Camacho2014},  potential solutions for the application of PCA to all referred Big Data problems are proposed.
{Velocity, volume and variety are addressed in} \cite{Wallace2016}{, where machine learning techniques are applied over a single variable to obtain signatures of known anomalies. The velocity and volume are handled by cloud computing over an Apache Spark cluster. The entropy of flows in network traffic is measured for anomaly detection in \cite{Wang2016}. {A Least-Mean-Square} adaptive filter is applied to obtain the correlation between the entropies and make better predictions. The velocity and volume are dealt with using an Apache Storm cluster.} The veracity is faced up in \cite{Krotofil2016}{, proposing an entropy-based anomaly detection system to identify attacks over the veracity of the system, such as data injection attacks. }%Finally, a PCA-based anomaly detection solution is introduced in} \cite{Camacho2014} {to solve the aforementioned Big Data problems}.

MBDA builds on previous works on MSNM and extends the latter for its application to Big Data. On the one hand, the Big Data functionality in the MEDA Toolbox \cite{Camacho201549,CSP} is employed to issue analytics on the  Big Data stream used to calibrate the anomaly detection model. This makes it possible to detect and isolate outliers that may affect the model performance. On the other hand, a new tool named the FCParser (https://github.com/josecamachop/FCParser) has been developed to work in perfect association with MSNM. This tool transforms raw data into analysis features and the other way round, so that we can extend the MSNM diagnosis to the original Big Data stream. This way, security professionals can use MBDA without any knowledge of multivariate analysis, something which was not possible with MSNM. This enhancement is expected to be very effective in reducing the time from detection to response in the face of cybersecurity incidents.

% SPC
\section{Multivariate Big Data Analysis: the 5 Steps}
\label{sec:statistical}

\begin{figure*}[tb]%[!ht]
	\centering
	\includegraphics[width=.8\textwidth]{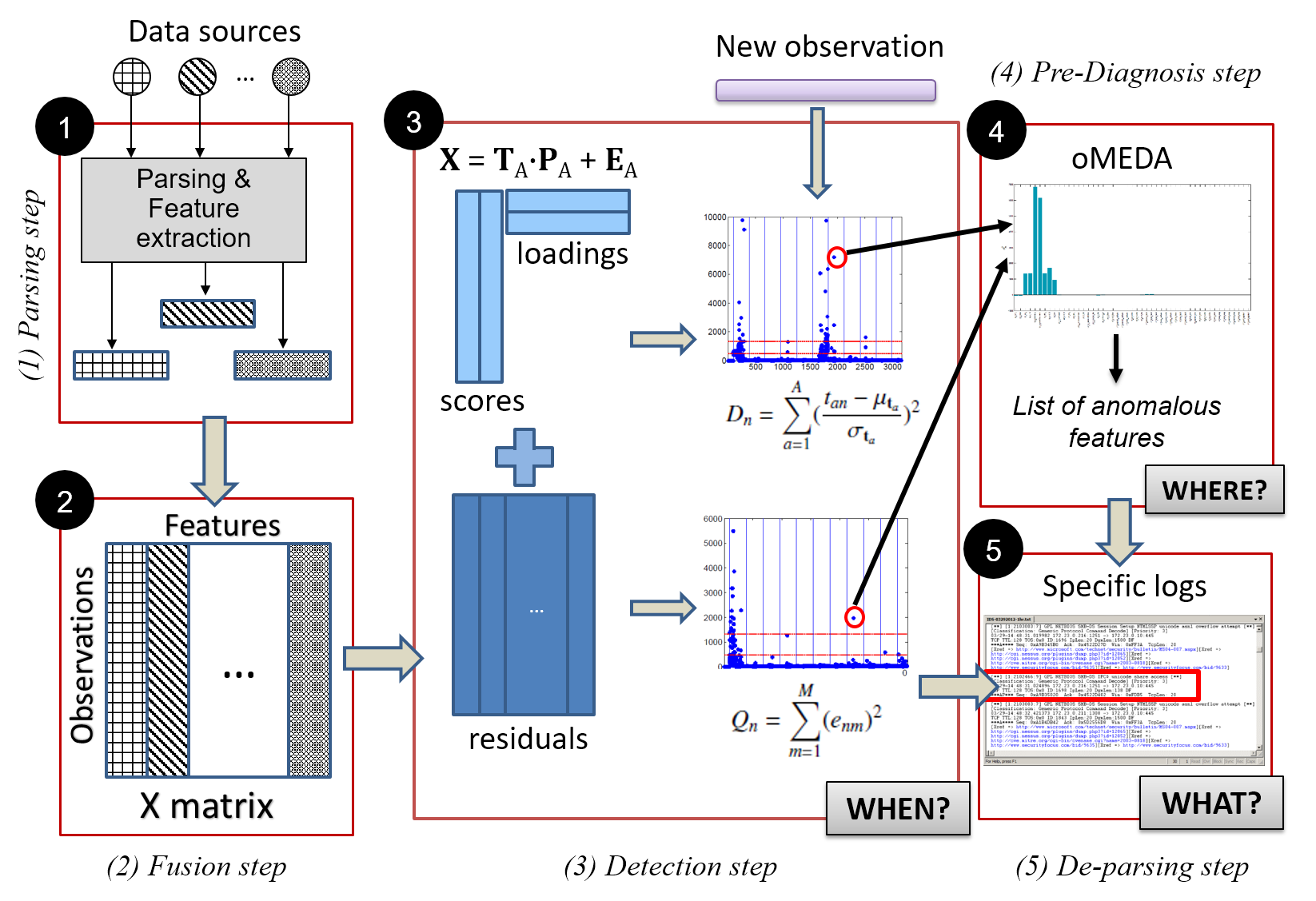}
	\caption{Illustration of the 5 steps of the proposed approach.}
	\label{fig:steps}
\end{figure*}

\begin{table*}[htb]
	\centering
	\caption{{Summary of the }five steps in MBDA.}
	\label{MBDA}

	\begin{tabular}{>{}l >{}l >{}l >{}l}

		\hline
		\hline
		\textbf{STEP} & \textbf{INPUT} & \textbf{OUTPUT} & \textbf{SOFTWARE}   \\
		\hline
		\hline
		1. Parsing & Raw data stream & Stream of features per source & FCParser   \\
		%\hline
		2. Fusion & Stream of features per source & Single feature stream & FCParser   \\
		%\hline
		3. Detection & Single feature stream & Timestamps for anomalies & MEDA Toolbox   \\
		%\hline
		4. Pre-diagnosis & Single feature stream  \& Timestamps for anomalies & Features for anomalies & MEDA Toolbox   \\
		%\hline
		5. De-parsing & Raw data stream \& Timestamps \& Features for anomalies & Raw logs for anomalies &FCParser    \\		
		\hline
		\hline
		
	\end{tabular}

\end{table*}

The MBDA approach consists of 5 steps:

\begin{itemize}
	\item[1)] Parsing: the raw data coming from structured and unstructured sources are transformed into quantitative features.
	
	\item[2)] Fusion: the features of the different sources of data are combined into a single data stream.
	
	\item[3)] Detection: anomalies are identified in time.
	
	\item[4)] Pre-diagnosis: the features associated with an anomaly are found.
	
	\item[5)] De-parsing: Using both detection and pre-diagnosis information, the original raw data records related to the anomalies are identified and presented to the analyst.
	
\end{itemize}

Notice the first three steps are equivalent to what it is commonly done in other machine learning methodologies. However, steps 4 and 5, which perform the diagnosis of the anomalies, are a main advantage of the present proposal. These steps are possible thanks to the white-box, exploratory characteristics of Principal Component Analysis (PCA) as the core of the MSNM approach. PCA is a linear model and as such it is easy to interpret in terms of the connection between anomalies and features, something much more complicated or not even possible in the non-linear machine learning variants, like for instance deep learning. The five steps, discussed in detail in what follows, are illustrated in Figure \ref{fig:steps} and summarized in Table \ref{MBDA}. { Note all these steps can be fully automatized.}

\subsection{Parsing}

The information captured from a network is usually presented in the form of system logs or network traces, and cannot be directly used to feed a typical tool for anomaly detection. Therefore, some sort of feature engineering and parsing needs to be done in order to generate quantitative features that can be used for data modeling. 

In the context of anomaly detection with PCA, Lakhina et \textit{al.} \cite{Lakhina2004} proposed the use of counters obtained from \textit{Netflow} records. In~\cite{Camacho2014}, we generalized this definition to consider several sources of data, proposing the feature-as-a-counter approach: essentially the combination of data counts with multivariate analysis. Each feature contains the number of times a given event takes place during a given time window. Examples of suitable features are the count of a given word in a log or the number of traffic flows with given destination port in a \textit{Netflow} file. This general definition makes it possible to integrate, in a suitable way, most sources of information into a model for anomaly detection. This parsing approach is also used for anomaly detection in X-pack \cite{Elastic}, the proprietary counterpart of the Elastic Stack, widely used for the analysis of Big Data streams. 

The parsing step is performed with the FCParser. For each feature we are interested on, a corresponding regular expression is defined in a configuration file. To obtain the specific values of the features in a given sampling interval, the corresponding regular expressions are run against the raw data of that interval, and the parser records the number of matches. 
{The selection of the specific features included in the parsing step is a manual process guided by expert knowledge of the data domain. Although it is desirable to have an automatic feature extraction process able to identify the relevant features to be considered in the parsing step, this is still open to future research. In the specific case study examples described in Sections 4 and 5 of this paper we describe how we have applied a manual selection of features.}

%So far, features are manually defined  from domain knowledge, but automatic feature extraction in the context of this technique is a very interesting future research. 

Among the five steps, the parsing step performs the most effective compression of the data, reason why it is instrumental for the application of MBDA to Big Data. Thus, the suitability of the MBDA approach for intrusion detection is grounded on the idea that we are capable of identifying security-related anomalies in the resulting parsed (compressed) information and of recovering the raw data related to such anomalies.  

\subsection{Fusion}

In the previous step, a set of features is defined for each different source of data. The sampling rate to derive the counts for each source may be chosen to be different, due to different dynamics of the sources or for convenience. Thus, to combine the features from different sources, these need to be stretched/compressed to a common sampling rate. Then, the features of the different sources are simply appended, yielding a unique stream of featured data of high dimensionality. The Fusion step is also done in the FCParser.

The combination of the feature-as-a-counter and the fusion procedure is specially suited for the subsequent multivariate analysis. It yields high dimensional feature vectors that need to be analysed with dimension reduction techniques, like PCA. Furthermore, counts and their correlation are easy to interpret. We will show that the diagnosis procedure in the 4th step, and the consequent 5th step, are benefited from the definition of a large number of features so as to better describe the anomaly taking place. This is the opposite in most anomaly detection schemes, where large dimensionality is seen as a concern, sometimes even referred to as a curse.

\subsection{Detection}

The core of MSNM is PCA. PCA is applied to data sets where $M$ variables or features are measured for $N$ observations. The aim of PCA is to find the subspace of maximum variance in the $M$-dimensional feature space.
The original features are linearly transformed into the Principal Components (PCs).
These are the eigenvectors of $\mathbf{X}^T \cdot \mathbf{X}$,
typically for mean centered $\mathbf{X}$ and sometimes also after auto-scaling, that is, normalizing variables to unit variance.

PCA follows the expression:
\begin{equation} \label{eq:PCAm}
\mathbf{X} = \mathbf{T}_{A} \cdot \mathbf{P}_{A}^{t} +
\mathbf{E}_A,
\end{equation}
where {$A$ is the number of PCs,} $\mathbf{T}_{A}$ is the $N \times A$ score matrix,
$\mathbf{P}_{A}$ is the $M \times A$ loading matrix and $\mathbf{E}_A$ is the $N \times M$ matrix of residuals. The columns in $\mathbf{P}_{A}$ are typically normalized to unit length vectors, so that the PCA transformation actually splits the variance in $\mathbf{X}$ into a structural part represented by the scores $\mathbf{T}_{A}$ and a residual part represented by $\mathbf{E}_{A}$. 

{For the detection of anomalies in} MSNM, a pair of statistics are defined: the D-statistic (D-st) or Hotelling's T2 statistic, computed from the scores, and the Q-statistic (Q-st), which compresses the residuals. Thus, using PCA and these statistics, we transform the problem of monitoring a highly dimensional multivariate stream of data $\mathbf{X}$ into the much simpler problem of monitoring a pair of statistics.

After a new sampling time interval, a new observation of the features $\mathbf{x}_{n}$ is computed. Subsequently, the corresponding  score vector is calculated as follows:
	
	\begin{equation} \label{eq:I_t}
	\mathbf{t}_{n} = \mathbf{x}_{n} \cdot \mathbf{P}_{A}
	\end{equation}
	
	\noindent where $\mathbf{t}_{n}$ is a $1 \times A$ vector with the corresponding scores, while
	
	\begin{equation} \label{eq:I_e}
	\mathbf{e}_{n} = \mathbf{x}_{n} - \mathbf{t}_{n} \cdot \mathbf{P}_{A}^{t}
	\end{equation}
	
	\noindent corresponds to the residuals. The D-st and the Q-st for observation $n$ can be computed from the following equations:
	
	{
		
		\begin{equation}
		D_{n} =  \mathbf{t}_{n} \cdot (\Sigma_T)^{-1} \cdot \mathbf{t}^t_{n} 
		\end{equation}
		
		\begin{equation}
		Q_{n} = \mathbf{e}_{n} \cdot  \mathbf{e}^t_{n}
		\end{equation}
		
		\noindent where $\Sigma_T$ represents the covariance matrix of the scores in the calibration data.} The values of $D_{n}$ and $Q_{n}$ are contrasted to the statistics of the calibration data to identify anomalies.

	A large percentage of the working effort of a CIRT is to analyze data related to potential incidents. If an efficient triaging method is available, more incidents can be detected by the same amount of personnel. As a result, in the practical application of MSNM to intrusion detection, we are more interested on the anomalies triaging rather than on the dichotomous distinction between what should be identified as anomaly and what should not. To combine the D-st and the Q-st into a single triaging score, we define here the \textit{Tscore} of observation $n$ according to the following equation: 
	
	\begin{equation} \label{eq:tscore}
	T_{n} = \alpha \cdot D_{n}/UCL^D_{.99} + (1- \alpha ) \cdot Q_{n}/UCL^Q_{.99}
	\end{equation}
	
	\noindent where $UCL^D_{.99}$ and $UCL^Q_{.99}$ are the upper control limits for the D-st and Q-st of the calibration data \cite{MSNM2015}, respectively, computed as $99\%$ percentiles, and $\alpha$ is weighting factor for the combination, which value  is discussed afterwards.
	
MSNM, following prior theory in the industrial field \cite{Ferrer2014}, establishes two phases { or modes of application. In the exploratory mode \cite{Lakhina2004} or {\it Phase I}, PCA is applied to a data block in order to find anomalies in that block. In the learning mode \cite{MSNM2015} or {\it Phase II}, PCA is calibrated from a data block to build a normality model, and then applied to new, incoming data, to find the anomalous events}. {\it Phase I} is devoted to network optimization, troubleshooting and situation awareness: essentially to detect, understand and solve any security-related problem and misconfiguration that was already affecting the network when the MSNM system was first deployed. These problems are the so-called special causes of variability in the argot of MSPC, because they induce unwanted variability in the data collected from the system under monitoring. \textit{Phase I} is carried out by detecting and diagnosing outliers in the data. The diagnosis is necessary to identify when there is a problem that needs to be solved by the technical personnel. For instance, if we identify an excess of blocked traffic in a gateway firewall, this can be the result of a cyber-attack  attempt that was correctly blocked by perimeter security measures, and therefore no further action needs to be applied. Alternatively, this can be the result of the firewall misconfiguration, which needs to be fixed. Following an iterative procedure in \textit{Phase I}, outliers are isolated and diagnosed, and corresponding problems identified and solved. When only non-relevant outliers remain, so that the network can be considered under normal operation conditions (NOC) or statistical control, we proceed to {\it Phase II}. In {\it Phase II}, the anomaly detector is used to identify anomalies in incoming data, typically in real time. The main idea beneath the definition of these two phases is that an anomaly detector should be developed only for a system under statistical control. Both phases in MSNM will be illustrated in the case studies of this paper.

Sometimes it is difficult to determine what should be understood as an outlier and what not, and therefore when to stop \textit{Phase I.} The practical approach we suggest to follow is to look for outliers in a bi-plot of the monitoring statistics or a barplot of the \textit{Tscore}. In those plots, outliers are easily identifiable. If no outlier is found, we start \textit{Phase II}. If otherwise any of the detected outliers reflects a real configuration/security problem, this needs to be solved and \textit{Phase I} re-started by measuring new traffic from the network. If no outlier identifies a practical problem, we still need to check whether the outliers pollute the PCA model or not. For that purpose, we can compare the amount of variance captured by each PC in the model with and without outliers. If this does not vary to a large extent, then we can proceed with \textit{Phase II}. Otherwise, we discard the outliers and re-start \textit{Phase I} using the remaining data.

It should be noted that the computation of the \textit{Tscore}, in particular the weighting parameter in eq. (\ref{eq:tscore}), differs in the two phases. In Phase I, the data we inspect for outliers is also the one used for model calibration. Therefore, outliers are intimately connected to the model variance, that is, outliers are expected to lay in the directions of high variance of the model. In this situation, we can set $\alpha$ to the percentage of variance captured by the model. Thus, we put more weight on the part that captures more variance, which depending on the case can be the model or the residual part. In \textit{Phase II}, however, this is not an adequate way of setting $\alpha$, since outliers are not part of the data used to fit the model and therefore they may separate from the rest of data in any potential direction of the space. Following \cite{432017}, in Phase II we use $\alpha$ equal to the ratio between the number of PCs $A$ and the number of variables $M$.  

The multivariate analysis in this paper is performed using the  MEDA toolbox \cite{Camacho201549}, which provides of a set of tools for multivariate anomaly detection. When data grows beyond a certain volume, the Big Data extension of the toolbox can be used. The basic principle of this extension is that multivariate models like PCA can be computed iteratively, in a way that is scalable to any data size and fully parallelizable. The loading
vectors of PCA can be identified using the eigendecomposition (ED)
of the cross-product matrix $\mathbf{X}^T \cdot
\mathbf{X}$, of dimension $M \times M$. This matrix can be computed in an iterative, incremental way as data is inputting the system, so that the number of rows in $\mathbf{X}$, $N$, is not a limitation any more \cite{Camacho2014}. To visualize the statistics of large numbers of observations, we use a clustering version of multivariate plots \cite{CSP}. The Big Data module of the MEDA Toolbox includes two computational cores: the iterative core and the exponentially weighted moving average (EWMA) core. Both algorithms solve the out-of-core computation of models and the corresponding clustering. The iterative core is used in this paper to compute the models.  

\subsection{Pre-Diagnosis}

{ Once an anomaly is signaled, a pre-diagnosis step is performed to identify the features associated with it. This information is very useful to make a first guess on the root causes of the anomaly. The contribution of the features to a given anomaly can be investigated with the contribution plots or similar tools. Among them, the most straightforward but effective approach is named Univariate-Squared (US) \cite{FUENTESGARCIA2018194}, and follows:
	
	\begin{equation} \label{eq:marta}
	d^2 = (\mathbf{x}_{n})^T \cdot |\mathbf{x}_{n}| 
	\end{equation}

\noindent Thus, anomalies are detected in the D-st and/or Q-st charts, and then the pre-diagnosis is performed with the US. The output of US is a $1 \times M$ vector  where each element contains the contribution of the corresponding feature to the anomaly under study. Those contributions with large magnitude, either positive or negative, are determined to be relevant. The computation of US for normal size data and Big Data is included in the MEDA toolbox.

\subsection{De-Parsing}

The last step of the MBDA approach is the extraction of the specific logs in the raw information that are associated with the anomaly. To accomplish that, we use both the information from the detection and the pre-diagnosis modules. The former provides the timestamps for the anomaly, which can be one or a set of consecutive sampling intervals. The latter provides the main features associated with the anomaly using eq. (\ref{eq:marta}). The de-parsing consist in reverting the parsing procedure selectively, obtaining, as a result, the raw logs related to the anomalies. To this end, the raw logs in the selected timestamps are matched against the pre-diagnosis features and sorted by the number of features they match. Depending on the data-set and the anomalies detected, the amount of information extracted in the de-parsing step could still be too large for visual inspection. For this reason, a user-defined threshold is set to limit the amount of retrieved data.

The FCParser is employed again in this step with the same configuration files used in the parsing step, where the regular expressions associated with the features were defined. In Algorithm \ref{deparsing}, the procedure of the de-parsing is detailed. Firstly, the algorithm goes through all the input files corresponding to the different data sources, looking for data records that occurred in the given timestamps ${T}$ and obtaining a selection of logs ${L}$. Then, the algorithm goes through all log lines in ${L}$ and assigns a score called ${f\!score}$. The ${f\!score}$ for a log line is the number of pre-diagnosis features ${(F)}$ that appears in that line. %The procedure is to identify how many features from ${F}$ could be extracted from a log in a similar fashion to the parsing step. 
This metric enables the algorithm to sort the log lines by relevance in an efficient way. Then, in a second loop, the extraction is performed. On each iteration, the algorithm extracts all the log lines that have ${f\!score}$ equal to  $N$, where ${N}$ is initialized to the number of features in ${F}$. In this manner, the log lines that contain all features ${F}$ are extracted first. If the number of log lines is not above the $threshold$, the number of features ${N}$ is reduced by one and the process is repeated. This is done for each data source until we reach the $threshold$ or $N$ reaches 0. 

The motivation for the de-parsing algorithm is that features identified in the pre-diagnosis step will likely be correlated, but do not necessary need to be present in all raw logs corresponding to an attack. Sometimes the attack is described by several types of logs that appear in the same time period. These logs will not contain all the features identified in the de-parsing, but a subset of them.

\begin{algorithm}
 \DontPrintSemicolon
 \KwIn{detection, pre-diagnosis}
 \KwOut{R}

$T~~ \leftarrow $~timestamps \tcp*{Timestamps from detection} 
$L~ \leftarrow $~select(T) \tcp*{Logs from timestamps} 
$F~ \leftarrow $~pre-diagnosis\tcp*{Anomalous features } 
$R~ \leftarrow $~[]\;
$N \leftarrow $~\#F \tcp*{Number of features to match}  
 
 \ForEach{logline in L}{
	$f\!score[logline] = $n\_feat(logline,F)  
					
		\tcp*{\# of anomalous features in the}
		\tcp*{logline}  
 }

 \While{$len(R) < threshold$}{
 	$R += extract(L,N,f\!score)$\; \tcp*{Extract lines with fscore=N}
 	$N \leftarrow N - 1$\; 
 }	

\Return R

 \caption{De-Parsing algorithm}
 \label{deparsing}
\end{algorithm}

{\subsection{Comparison of MBDA 5-steps with State-of-the-Art Methodologies}}
{In this section, we compare the main state-of-the-art anomaly detection methodologies discussed in Section~\ref{sec:related} with our proposal, taking the 5-steps as a comparison framework. Table~\ref{5stepsCMP} relates the aforementioned contributions to the 5 steps. A checkmark is shown for those approaches that consider the corresponding step. If the checkmark is into parenthesis, only  partial functionality of such step is implemented. }%the results are not correct,  either the step does not match with the description in this work. }

\begin{table*}[htb]
	\centering
	\caption{{State-of-the-art anomaly detection methodologies compared to the MBDA 5 steps.}}
	\label{5stepsCMP}
	\begin{tabular}{>{}l  >{}c >{}c | >{}c >{}c >{}c >{}c >{}c}
		\hline
		\hline
		\textbf{Methodology} & \textbf{Years} & \textbf{Ref.} & \textbf{Parsing} & \textbf{Fusion} & \textbf{Detection} & \textbf{Pre-diagnosis} & \textbf{De-parsing} \\
		\hline
		PCA-based I & 2003  & \cite{Kanaoka03} & \checkmark & \xmark &\checkmark &\xmark & \xmark     \\
		%-& \cite{Mikel2017Auth} & \checkmark & \checkmark & \checkmark &\xmark & \xmark     \\
		SVM & 2003--2015 & \cite{Heller2003,HONG2015} & (\checkmark) & (\checkmark) & \checkmark &  \xmark &  \xmark   \\
		PCA-based II & 2004--2018 & \cite{Lakhina2004, Delimargas2014,Callegari2014,Xia2018} & \checkmark  & \checkmark & \checkmark & (\checkmark) &   \xmark   \\
		Clustering & 2009--2016 & \cite{Gupta2016, Gonsalves2015,Hurst2014,Rathore2016,Wallace2016,Xu2009, Yen2013,Therdphapiyanak2013a} & \checkmark  &\checkmark &\checkmark & \xmark &  \xmark    \\
		HASHDOOP & 2014&  \cite{Fontugne2014} & \checkmark  & \checkmark& \checkmark& \xmark & \xmark     \\
		Entropy Calculation I & 2015 & \cite{Krotofil2016} & (\checkmark) & \xmark  &\checkmark  &\xmark & \xmark     \\
		Entropy Calculation II - TADOOP & 2015 & \cite{Tian2015}& \checkmark & \xmark &\checkmark & \xmark &  \xmark    \\
		Entropy Calculation III & 2016 & \cite{Wang2016} & \checkmark & \checkmark & \checkmark & \xmark& \xmark     \\
		MSNM & 2016--2017 & \cite{MSNMSP17,432017} & \checkmark & \checkmark &  \checkmark & \checkmark  & \xmark  \\
		MBDA 5-steps & 2019 & & \checkmark & \checkmark &  \checkmark & \checkmark  & \checkmark \\		
		\hline
		\hline
	\end{tabular}
\end{table*}

{Methodologies based on traditional machine learning \cite{Heller2003,Hadziosmanovic2012,HONG2015} typically perform steps 1~to~3, not dealing with pre-diagnosis (step 4) or de-parsing (step 5). The same applies for machine learning techniques focused in Big Data anomaly detection \cite{Gupta2016, Gonsalves2015,Hurst2014,Rathore2016,Wallace2016,Xu2009, Yen2013,Therdphapiyanak2013a}. Furthermore, some of the approaches that combine traditional methodologies with Big Data tools \cite{Tian2015, Krotofil2016}, or multivariate techniques like \cite{Om12,Kanaoka03} only mention step 1 (feature engineering) and step 3 (detection).}
	
{The PCA techniques based in Lakhina's work \cite{Lakhina2004, Fernandes2016,Aiello2016, Jiang2015,Chen2016,Xia2018} are more complete, in the sense that they consider steps 1~to~4. Still, the diagnosing method is limited in those methods and improved in MSNM \cite{MSNMSP17,FUENTESGARCIA2018194}. Yet, in MSNM works the de-parsing step was not considered.}

\section{Case Study I: VAST Challenge}
\label{sec:case2}
% CASE STUDY
\subsection{Experimental Framework}
The data set comes from the VAST 2012 2nd mini challenge \cite{VAST2012}, a publicly available dataset used for the IEEE Scientific Visualization Conference of the year 2012. The VAST-MC2 presents a corporate network scenario where security incidents occur during two days. In particular, some of the staff report unwanted messages and a non-legitimate anti-virus program appearing on their monitors. Also, their systems seem to be running more slowly than normal. In summary, a forensics operation is required to discover the most relevant security events and their root causes. As the challenge is from the past, we know the solution beforehand: a botnet infected the network and attacked the DNS servers,
causing performance issues and the infection of workstations with adware.

This case of study serves to illustrate \textit{Phase I} {(exploratory mode)} for anomaly detection in an environment with disparate security data sources. \textit{Phase II} is not feasible because the network is already infected and data under NOC is not available. The main objective of \textit{Phase I} is network troubleshooting and optimization, detecting anomalies that leads to problems and misconfiguration.

The network infrastructure for the challenge consists of approximately 4,000 workstations and  1,000 servers that operate 24 hours a day. The intra-network nodes are in the IP range of 172.x.x.x and IPs from other ranges are considered external. The data provided with the VAST 2012 mini challenge 2 consist of Cisco ASA firewall logs including a total of 23,711,341 data records, and intrusion detection system logs including 35,948 data records. The dataset details are available at \cite{VAST2012}.

Reproducibility of results in this case study is possible by downloading the virtual machine at https://nesg.ugr.es/veritas/index.php/mbda

\iffalse

\noindent
\begin{figure*}[!ht]
	\centering
	\includegraphics[width=.85\textwidth]{figuras/VAST/VAST_topology.jpg}
	\caption{Network topology VAST 2012 MC2 \cite{VAST2012}}
	\label{fig:VAST_topo}
\end{figure*}

\fi

\subsection{Application of the 5-steps methodology}

\textit{{\textbf{Parsing \& Fusion steps.}}} First of all, using the FCParser tool, the data from the two data sources is parsed and fused into observations for analysis. Firewall and IDS logs are semi-structured data sources because the format of the different data entries is not fixed. Thus, this experiment illustrates how the FCParser can be used to deal with different kinds of sources of information with uneven formats.

From the raw data, 265 features have been designed, 122 for firewall logs and 143 for IDS logs. In Tables \ref{featFW} and \ref{featIDS} an overview of the features can be found. We decided to use 1 minute time interval for each observation, obtaining, as a result, a 2345 x 265 matrix of parsed data. This reduces the 4.2 GB of raw data to less than 2.5 MB, showing that this step is effectively used as a compression step.

\begin{table}[htb]
\centering
\caption{Firewall features overview }
\label{featFW}
\begin{tabular}{c c c}

\hline
\hline
Number of features & Data considered in the features\\
\hline

6 & Source and destination IPs \\
84 & Source and destination ports  \\
13 & ASA Messages \\
2 & Protocol \\
8 & Syslog priority \\
7 & Action \\
2 & Direction of conection \\

\hline
\hline
\end{tabular}
\end{table}

\begin{table}[htb]
\centering
\caption{IDS features overview}
\label{featIDS}
\begin{tabular}{c c c}

\hline
\hline
Number of features & Data considered in the features \\
\hline

6 & Source and destination IPs \\
84 & Source and destination ports \\
4 & Snort priority \\
33 & Snort Classification \\
10 & Snort event description \\
6 & IP headers \\

\hline
\hline
\end{tabular}
\end{table}

\textit{{\textbf{Detection step.} }}The  relevance of the different features in terms of security is heterogeneous. For example, a connection to the port 80 found in the firewall logs is not as important as a SSH brute force attack detected in the IDS logs. For this reason, to establish levels of severity, each feature is assigned a weight from 1 to 10. This weight is multiplied to the data after auto-scaling, so that the higher the weight, the higher the relevance of the variable in the model.

Once the parsed data is prepared, the MSNM \textit{Phase I} analysis is performed. For that, the PCA model is obtained and the D-st and Q-st are computed for each observation. Because the data volume is reduced, we did not employ the Big Data functionality in the MEDA Toolbox, which will be illustrated in the other case study. With the monitoring statistics, the \textit{Tscore} in Eq. (\ref{eq:tscore}) is computed for observation triaging. Figure \ref{fig:VAST_Tscore} shows the \textit{Tscore} values for the complete two day interval. The 5 observations that stand out the most are selected for further analysis: 369, 370, 1413, 389, 384.

\textit{{\textbf{Pre-diagnosis step.} }}We use the US tool to extract information about the selected anomalies following Eq. (\ref{eq:marta}). This procedure yields a subgroup of features that are related to the given anomalies providing insight into those events. The US scores with the highest absolute magnitude highlight the main variables related to an anomaly. A positive score for a variable means that the anomaly presents an unexpectedly high value of that variable, while a negative score means right the opposite: the value of the variable is lower than expected. Many times, US scores are informative enough to avoid the need to look at the original raw data, that is, to perform the deparsing step. This makes diagnosis faster and, in turn, response to security incidents faster. 

In Figure \ref{fig:oMEDA_369}, the US plot of the observation number 369, corresponding to timestamp 2012-04-06 00:04, is shown. The features with values that stand out from the others are selected. Table \ref{table:diag} shows the features associated to each of the five observations identified in the detection. Those results demonstrate that MBDA makes the most of the combination of the two data sources, as the US yields feafures from both sources, for example, for observations 369 and 370.

\noindent
\begin{figure}[!bt]
  \centering
  \hspace*{-1.5cm}\includegraphics[width=.61\textwidth]{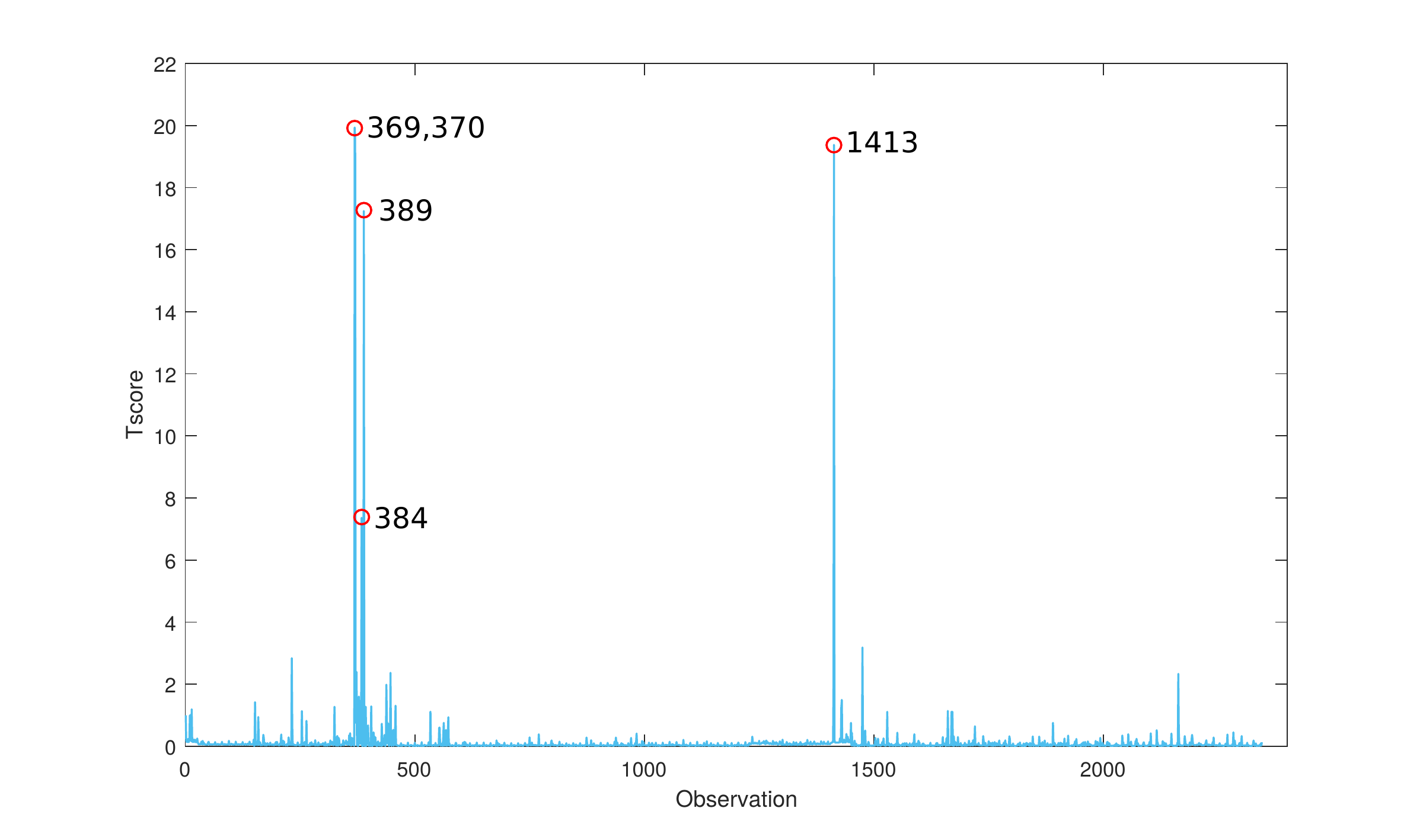}
  \caption{Tscore values for each observation.}
  \label{fig:VAST_Tscore}
\end{figure}

\noindent
\begin{figure}[!bt]
	\centering
	\hspace*{-1.5cm}\includegraphics[width=.61\textwidth]{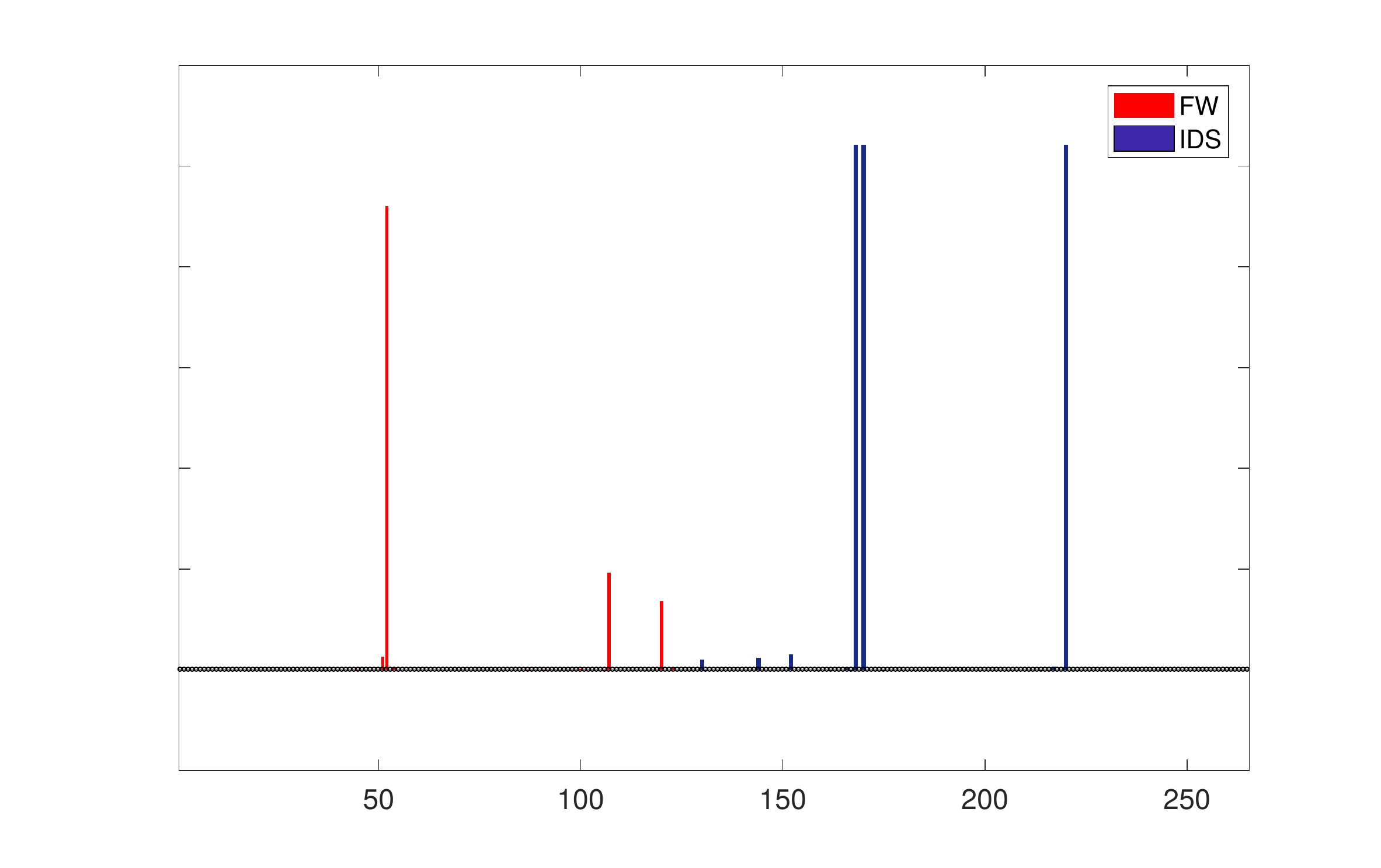}
	\caption{US plot for the observation 369.}
	\label{fig:oMEDA_369}
\end{figure}

\begin{table*}[htb]
	\centering
	\caption{VAST2102: Anomaly Diagnosis.}
	\label{table:diag}
	\begin{tabular}{c c c c c c c c}
		
		\hline
		\hline
		Index & Timestamps & Tscore & Sources & Features selected & logs & log types & Deparsing (Interpretation) \\
		\hline

		&             &          &           &  fw\_dport\_telnet ids\_ssh\_scan      &      &     &  Data exfiltration attempt  \\
		369  & 06/04 00:04 &  19.94   & ids \& fw &  ids\_ssh\_scan\_outbound              &  32  &  3  &  by SSH and Telnet, the latter \\
		&             &          &           &  ids\_dport\_ssh                       &      &     &  blocked by access lists.    \\ \hline
		
		&             &          &           &  fw\_dport\_snmp fw\_denyacl           &      &     &                 \\
		&             &          &           &  ids\_dport\_snmp ids\_snmp\_req       &      &     &   Attempted information            \\
		370  & 06/04 00:05 &   19.52   & ids \& fw &  ids\_successful-recon-limited         &   71 &  4  &   leak using the SNMP protocol.     \\
		&             &          &           &  fw\_error  ids\_brute\_force          &      &     &              \\ 
		&			 &			&           &  ids\_attempted-recon	                 &      &     &  \\ \hline
		
		384  & 06/04 00:19 &   7.36  & ids       &  ids\_scanbehav                        &  2   &  1  &  Scan for windows RDP  \\
		&             &          &           &                                        &      &     &  ports for vulnerabilities. \\ \hline
		
		&             &          &           & ids\_attempted-recon                   &      &     &  Scan ports in range      \\
		389  & 06/04 00:24 &  17.25   & ids       & ids\_successful-recon-limited          &  91  &  1  &  5900-5920 looking               \\ 
		&             &          &           & ids\_prio2  ids\_vnc\_scan             &      &     &  for vulnerabilities.          \\ \hline

		1413 & 06/04 17:28 &  19.38   &  ids      &    ids\_policy-violation               & 254  &  1  &  DNS server attack from       \\
		&             &          &           &  ids\_prio1 ids\_dns\_update           &      &      &  multiple systems.    \\ \hline

		\hline
		\hline
	\end{tabular}
\end{table*}

\textit{{\textbf{De-parsing step.} }}Finally, from the timestamps of the events and the features related to anomalies, the specific raw logs associated with the detected anomalies  are extracted. This is achieved %with the de-parsing process 
using the FCParser tool. The last three columns in Table \ref{table:diag} summarize the results of de-parsing the 5 anomalies from the anomaly detection step. To sum up, a total of 450 log entries are retrieved, representing a  {0.0019\%} of the logs in the original data. However, there are only 10 different log types, representing only {0.00004\%} of all the input data, which makes the analyst interpretation of the results straightforward. In the following this interpretation is discussed.

At 06/04 00:04, numerous attempts of data exfiltration by Telnet and SSH were discovered. The firewall blocked the Telnet connections, however, the SSH connections passed through. 

At 06/04 00:05, an information leakage carried out using SNMP was detected by the IDS. The deparser also shows that those requests were blocked by the firewall.

At 06/04 00:19 and 06/04 00:24 we can see multiple vulnerability scans targeting remote desktop services like VNC and RDP. In particular, the detection of just two logs related to RPD shows that the MBDA approach is especially sensitive to events which are not common in the traffic, being an adequate tool to identify those events as soon as they show up.

At 06/04 17:28 the logs extracted from the deparser spotlight a coordinated attack from multiple infected systems targeting the DNS server at 172.23.0.10. If the DNS server is compromised, that would be the reason for the ad-ware and malicious anti-virus programs that appeared on the systems. %At 06/04 18:30 the attacker tried to exfiltrate data by FTP from multiple infected systems but, once again, this traffic was blocked by the firewall. 

\textit{{\textbf{Summary. }}}This example shows that with the proposed approach and one single iteration of the \textit{Phase I} analysis, we can extract most of the relevant information on the compromise of the network in the present example. Just 10 log types are enough to shed light on what has occurred on the most anomalous events, drastically reducing the time from detection to response and consequently, the cost derived from those events.

\section{Case Study II: ISP Network}
\label{sec:case}
% CASE STUDY
\subsection{Experimental Framework}

\begin{table}[t]
	\caption{Characteristics of the calibration and the test sets.}
	\label{tab:setsFeatures}
	\centerline{
		\small{
			\begin{tabular}{l c c}
				\hline \\[-1.5ex]
				{\bf Feature} & {\bf Calibration} & {\bf Test} \\[0.5ex]
				\hline \\[-1.5ex]
				Capture start & 10:47h 03/18/2016 & 13:38h 07/27/2016 \\
				Capture end & 18:27h 06/26/2016 & 09:27h 08/29/2016\\
				Attacks start & N/A & 00:00h 07/28/2016 \\
				Attacks end & N/A  &  12:00h 08/09/2016\\
				Number of files & 17 & 6 \\
				Size (compressed) & 181GB & 55GB \\ 
				\# Connections & $\approx$ 13,000M & $\approx$ 3,900M\\[0.5ex]
				\hline
			\end{tabular}
	}}
	
\end{table}

In this section, we present the 5-steps methodology applied to a real network scenario, taking as input the UGR'16 dataset. For the sake of completeness, we provide a brief description of the dataset in what follows, although all the details can be found in~\cite{Dataset}.  

We collected this data on a real network of a Tier~3 ISP. The services provided by the ISP are mainly virtualization and hosting. To a shorter extent, the ISP also provides common Internet access services. 

 {\it Netflow} sensors were deployed in the {border routers of the} network to collect  traffic corresponding to its normal operation. In addition, a total of 25 virtual machines were deployed in order to perform controlled malicious activities. Some of these virtual machines were used to launch a number of specific attacks over time against the rest, which acted as the victims of the attacks.
 %The obvious ethical reason for using virtual machines as victims of  the controlled attacks is to avoid negative effects on the normal operation of the legitimate services. Both the attackers and the victims are inside the ISP infrastructure to avoid the potential detection and blocking of the attacks by other intermediate ISPs. 
 The ISP personnel was aware and collaborated in the experiment.

We obtained two sets of data. A \textit{calibration set} with more than three months of traffic and a \textit{test set} of appr. one month. The main characteristics of both sets are shown in Table~\ref{tab:setsFeatures}. Attacks were generated during the test set time period, in intervals of 2 hours and alternating with legitimate traffic for 12 consecutive days. {This example illustrates the application of the proposal through \textit{Phases I \& II} (exploratory and learning mode).}

\subsection{Application of the 5-steps methodology}

%In this case study, we perform in detail the Phase I analysis of the \textit{calibration set} and then we proceed with the Phase II using the \textit{test data}.

\textit{{\textbf{Parsing step. }}}In the first step, data from the calibration and test datasets are parsed into $M$-dimensional vectors (observations) representing time intervals of 1 minute. In particular, we defined a set of $M$=138 network-related features, corresponding to {11} different \textit{Netflow} variables as shown in Table \ref{tab:vars}. The 1 minute interval leads to app. 144K observations in the calibration data, with an storage volume of app. 80 MB. { A single machine with 16 cores running the FCParser parses 100M connection logs in 3 hours and 16 seconds, and computations can be fully parallelized. Thus, the complete computation of 13,000M connection logs in our calibration data set can be done in app. 16 days in a single, 16 core, machine, or in little more than 6 hours in a large-size parallel cluster of 1000 cores. A main limitation to speed-up the parsing is the use of python as the programing language of the FCParser. Regular expressions in python are 
much slower than in other programming technologies.	To check an alternative technology in order to reduce processing times, we also programmed  an ad-hoc parser in C, which was observed to be 100-times faster. The latter defines the regular expressions corresponding to the features at code level, unlike the FCParser, where the selection of features is done at configuration level, i.e. in configuration files. The definition of the features in configuration files is slower, but simplifies the application of the 5-steps methodology in a new problem and the re-definition/modification of features.}

{ Thanks to the compression ability of the first step, the rest of steps can be performed in a regular computer.}

\textit{{\textbf{Fusion step.} }}In this example we work on a single source of data: \textit{Netflow}. Therefore, no data fusion is necessary.

\begin{table}[t]
	\caption{Overview of Features in the second case study.}
	\label{tab:vars}
	\centerline{
		\small{
			\begin{tabular}{|l|l|}
				\hline
				{\bf Variable} & {\bf \#features $\rightarrow$ values} \\
				\hline
				Source IP & 2 $\rightarrow$  {\it public, private} \\
				Destination IP & 2 $\rightarrow$  {\it public, private} \\
				Source port & 50 $\rightarrow$  {\it specific services, Other} \\
				Destination port & 50 $\rightarrow$  {\it specific services, Other} \\
				Protocol & 5 $\rightarrow$  {\it TCP, UDP, ICMP, IGMP, Other} \\
				Flags & 6 $\rightarrow$  {\it A, S, F, R, P, U} \\
				ToS & 3 $\rightarrow$  {\it 0, 192, Other} \\
				\# Packets in & 5 $\rightarrow$  {\it very low, low, medium, high, very high} \\
				\# Packets out & 5 $\rightarrow$  {\it very low, low, medium, high, very high} \\
				\# Bytes in & 5 $\rightarrow$  {\it very low, low, medium, high, very high} \\
				\# Bytes out & 5 $\rightarrow$  {\it very low, low, medium, high, very high} \\
				\hline
			\end{tabular}
	}}
\end{table}

{\textit{\textbf{Phase I. Detection step.}}} In this case study, we first detail the Phase I analysis of the \textit{calibration set}, and then we proceed with the Phase II using the \textit{test data}. In what follows, steps 3 to 5 are illustrated for both phases.}

While the previous steps led to a large compression of the information (from 181GB to 80 MB), the numbers in terms of observations are still too large for the direct computation of the PCA model. Thus, in this example, the Big Data module of the MEDA Toolbox is employed.

%\begin{figure*}[!tb]%[!ht]
%	\centering\includegraphics[width=0.6\textwidth]{./figuras/ACSAC/MSPC2wMC.eps}
%	\caption{Compressed Multivariate Statistical Network Monitoring plot for calibration data: scatter plot of the Q-st vs the D-st.}
%	\label{fig:ACSAC_MSPC2w}
%\end{figure*} 

\begin{figure}[!ht]%[!ht]
	\centering\includegraphics[width=0.5\textwidth]{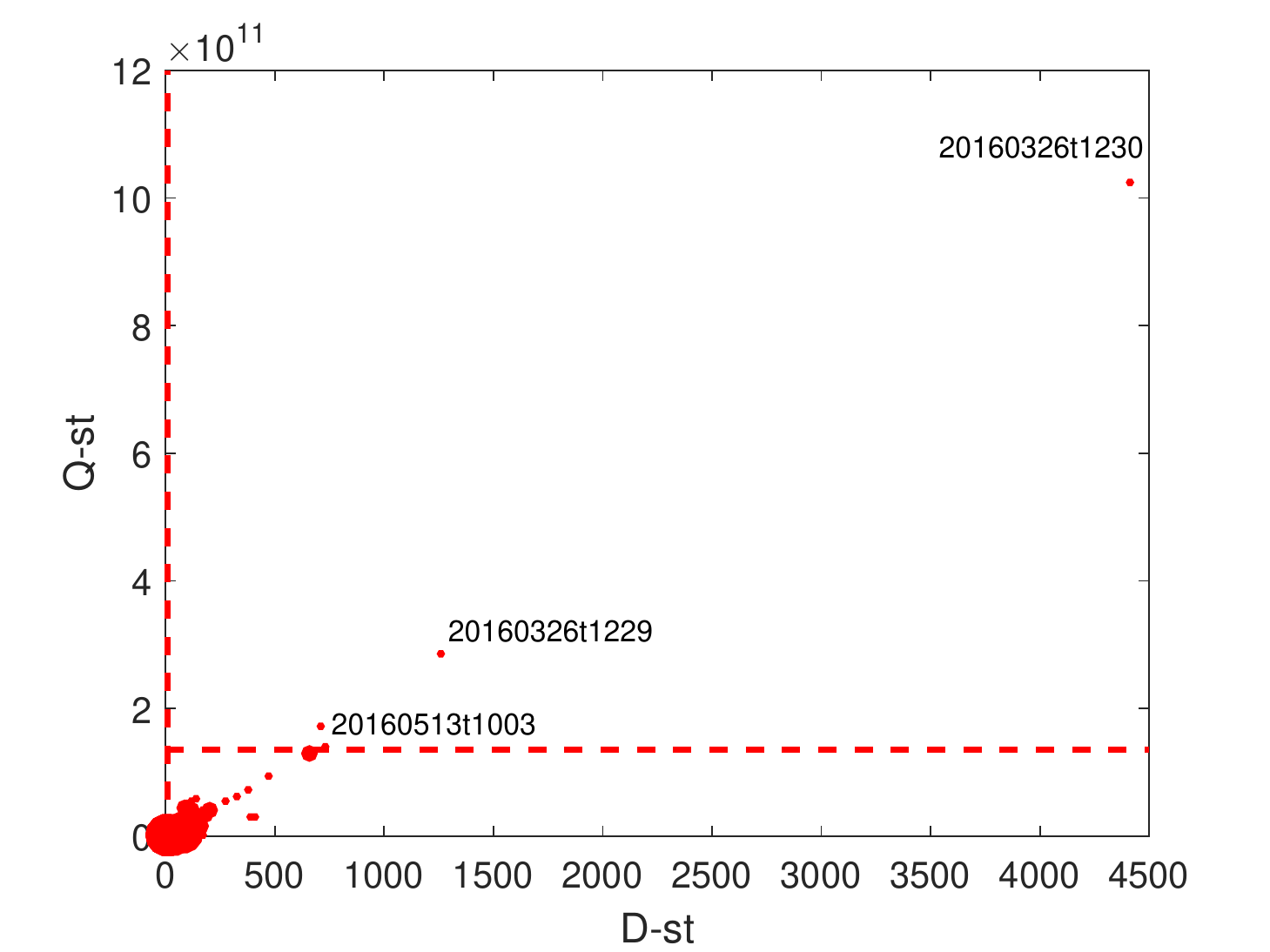}
	\caption{Compressed MSNM plot for the calibration set: scatter plot of the Q-st vs the D-st. Outliers are labeled with their timestamps.}
	\label{fig:MSPC2w}
\end{figure}

The compressed MSNM plot for Phase I analysis is shown in Figure \ref{fig:MSPC2w}. The plot is a scatter plot of D-st values vs Q-st values, which is an alternative to the Tscore bar plot. Thanks to the functionality in the Big Data module of the MEDA Toolbox, the 144K observations were clustered in 100 clusters to improve visualization. The size of the markers depends on the {multiplicity} of each cluster, that is, the number of original observations in it. This is particularly useful in anomaly detection, where outliers typically show up as individual or very small clusters. Control limits containing the $99\%$ percentiles of the calibration sample are shown in the plot to facilitate the identification of outliers. Labels in the plot display the timestamp of several individual observations. We can see that there is an excursion from normal traffic at day 2016-03-26 between 12:29 and 12:30, and another at 10:03 of 2016-05-13. 

\begin{table}[!tb]
	\centering
	\caption{Most relevant variables in the US pre-diagnosis for outliers 20160326t1229-20160326t1230 and 20160513t1003.}
	\label{tab:t20160513t1003}
	\begin{tabular}{c}
		\hline
		\hline 
		Time interval 20160326t1229-20160326t1230 \\
		\hline
		`npackets\_verylow'    `srctos\_zero'    `tcpflags\_ACK' \\
		    `dstip\_public' `srcip\_public'  `protocol\_udp'  \\
		   `nbytes\_medium'    `dport\_register'  
		`sport\_register' \\
		
		\hline
		\hline
		Time interval 20160513t1003 \\
		\hline
		   `dstip\_public'    `srcip\_public'    
		`tcpflags\_ACK'   \\  `protocol\_tcp'    `tcpflags\_SYN' 
		`npackets\_verylow' \\   `nbytes\_verylow'  `srctos\_zero'  
		`dport\_reserved'  \\`dport\_smtp'    `sport\_reserved'   
		`srctos\_other'  \\  `tcpflags\_RST'    `sport\_smtp' \\
		
		\hline
		\hline
	\end{tabular}
\end{table}

{\textit{\textbf{Phase I. Pre-diagnosis step.}}} In Table~\ref{tab:t20160513t1003}, we show the features with highest magnitude according to the US for the two outliers identified. For day 2016-03-26, the pre-diagnosis information in the Table is too general, and we can search for more detail with the MBDA 5th step. 
	
{\textit{\textbf{Phase I. De-parsing step.}}} Since the FCParser cannot handle binary data in its current version, we will manually perform the de-parsing step with \textit{nfdump} commands. We run the following command, extracted from the timestamps and the definition of the features:\\
\\
\noindent \texttt{PROMPT\$ nfdump -r <nfcapd.file> -t 2016/03/26.12:29:00-2016/03/26.12:30:59  `(packets <4) and (bytes > 1000 and bytes < 10001) and (dst port > 1024 and dst port < 49151) and (src port > 1024 and src port < 49151) and (flags A or proto udp)' }\\

\noindent where we intentionally combined `tcpflags\_ACK' and `protocol\_udp' with an 'or' junction, since they are obviously contradictory. 

The query provided 436K flows out of the total $\approx$ 1.2M flows in the two minute interval, which corresponds to 0.00003\% of the total size of the calibration data (a needle in a haystack). From those, all but 80 flows were related to the anomaly (99.98\% of specificity). The flows reveal a scanning operation from an internal machine (42.219.156.231) to all the 65535 possible UDP ports at machine with anonymized IP 62.151.13.8. Further inspecting among the two IPs uncovered a total of 1M flows.

For day 2016-05-13, the pre-diagnosis results show there was an unexpected excess of very short SMTP connections. To proceed with step 5, we run the following command:
\\\\
\noindent \texttt{PROMPT\$ nfdump -r <nfcapd.file> -t 2016/05/13.12:29:00-2016/05/13.12:30:59 `packets < 4 and bytes < 151 and dst port = 25 and (flags S or flags A or flags R)'}\\ 
 
 \noindent where again flags are combined with 'ors'.
 
The query yielded a total of app. 125K flows out of the 400K flows in that minute interval, from which all but 15 were related to the anomaly (99.99\% of specificity). To interpret the result, the flows only need to be aggregated by origin IP, leading just to 4 public IPs following the pattern of a SPAM campaign. The ISP IT team confirmed this was a client that hired virtual machines during a period. The hired IPs finished in the YAHOO blacklists, and that is probably the reason why the client stopped hiring the virtual machines. This misuse behavior was repeated throughout the capture interval.

According to the MSNM philosophy discussed before, the problems identified in the data need to be solved and further data need to be collected to continue Phase I. Since we are working on a previous capture, we cannot collect new data. However, to continue the analysis an alternative is to filter out the netflow traces isolating the attacks, and repeat Phase I from the rest of data. After this second iteration, we stopped Phase I to proceed with Phase II.

\begin{figure}[tb]%[!ht]
	\centering
	\includegraphics[width=.45\textwidth]{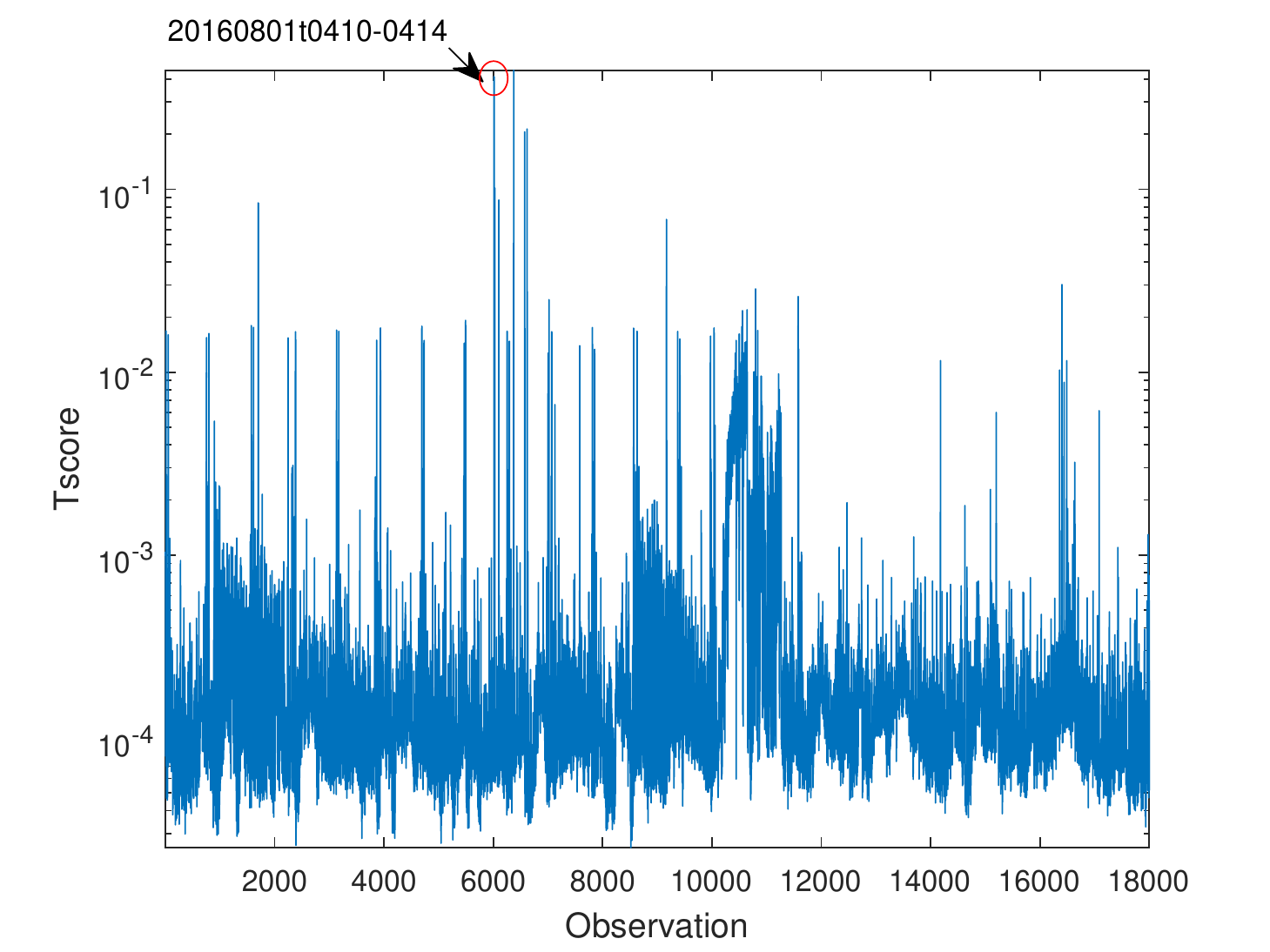}
	\caption{\textit{Tscores} per test observation: higher values represent identified anomalies. }
	\label{fig:data31}
\end{figure}

{\textit{\textbf{Phase II. Detection step.}}} Once we finish Phase I, we have a definitive normality model of the traffic and can proceed to analyze incoming data in real time. This is exemplified using the test data. The Tscore plot of the test data is shown in Fig. \ref{fig:data31}. While a compressed MSNM plot like the one shown before could be computed for this data, the toolbox only includes its computation for calibration data, that is, data that is used to generate the model. Still, the Tscore provides the identification of the outliers. We can see there are outliers above the regular spikes, the latter that correspond to synthetic attack periods.
%In Figure~\ref{fig:data31}, we represent the values of the \textit{Tscores} obtained in the first period of the \textit{test set}.%, when synthetic anomalies (attacks) were introduced
%. The higher the \textit{Tscore} for a given sampling time, the more confidence to determine there is an anomaly in that instant. %There is a high degree of coincidence among the two methods, detecting the same or close time points. 
%Note the periodic spikes that correspond to the synthetic anomalies introduced in the dataset. 
%{In the figure,} dark (red) circles represent top-most scores in each detector. {Among them, those that correspond to induced attacks are also identified by dark (black) squares}. It is specially remarkable that very few of top-most detections correspond to induced attacks, but to true attacks in the background traffic \cite{Dataset}.
The most relevant anomaly corresponds to the interval 20160801t0410 to 20160801t0414, annotated in Figure~\ref{fig:data31}. 

%Note this type of diagnosis is not possible in OCSVM. We will focus {on} the two  intervals labeled in Figure \ref{fig:data31}: time intervals from 20160801t0410 to 20160801t0414 and from 20160806t2039 to 20160807t0559, signaled as anomalous by the {two} methods but which do not correspond to synthetic attacks. { The list of variables identified by oMEDA for these examples are listed in Table \ref{tab:t20160730t0452}. }

%\begin{figure}[tb]%[!ht]
%	\centering
%	\includegraphics[width=0.4\textwidth]{./figuras/t20160730t0452_2w.eps}
%	\caption{oMEDA plot for time point 20160730t0452.}	\label{fig:t20160730t0452}
%\end{figure}

\begin{table}[]
	\centering
	\caption{Variables in order of relevance according to the oMEDA diagnosis for the selected anomaly in the \textit{test set}.}
	\label{tab:omedaLabelsTestSet}
	\begin{tabular}{c}
		\hline
		\hline
		Time interval 20160801t0410-0414 \\
		\hline 
		`srctos\_other'    `sport\_register'    `nbytes\_low'\\    `dport\_register'    `npackets\_verylow' \\
		`protocol\_udp'    `srcip\_public'    `dstip\_public'\\    `tcpflags\_ACK'    `label\_background'\\
		`srctos\_zero' \\	
		\hline
		\hline
	\end{tabular}
\end{table}

{\textit{\textbf{Phase II. Pre-diagnosis step.}}} The US pre-diagnosis of this short interval %from 20160801t0410 to 20160801t0414 
shows an increase of ACK packets and very short connections using UDP (see relevant features in Table~\ref{tab:omedaLabelsTestSet}), a pattern that resembles that of the first anomaly in the calibration data set, diagnosed as a scanning attack. 

{\textit{\textbf{Phase II. De-parsing step.}}} Proceeding with the 5th step, we find a single IP from Germany creating 800K connections (0.0002\% of the total) from origin ports 5061, 5062, 5066 and 5069. The destinations are 4097 different hosts in 16 different subnets with /24 mask. Each host is scanned through ports 6000-6060. The whole time interval contained more than 1M flows, and the query accurately identified the connections of interest plus a limited number of less that 1000 connections (specificity of 99.87\%). {According to the IT staff,} the event seems to be a malware driven scanning, due to this specific pattern of connection.

{\textit{\textbf{Summary. }} This example illustrates the application of the 5-step methodology in a real data set, showing the viability of both \textit{Phase I} and \textit{Phase II} analyses. In both situations, the methodology could identify real attacks with a level of specificity higher than  99\% and providing a reduced set of logs for analysis. This experience also showed that the computation of the 5-steps can be orders of magnitude faster using parallelization hardware and/or a fast processing language. Permanent deployments of this technique should take into account this finding.}

%\noindent \texttt{PROMPT\$ nfdump -r <nfcapd.file> -t 2016/08/01.04:10:00-2016/08/01.04:14:59  `(packets <4) and (bytes > 150 and bytes < 1001) and (dst port > 1024 and dst port < 49151) and (src port > 1024 and src port < 49151) and (flags A or proto udp)' }\\.  

% CONCLUSION
\section{Discussion and Conclusion}
\label{sec:conclusion}

In this paper, we propose a Big Data anomaly detection methodology named Multivariate Big Data Analysis (MBDA). The approach is based on 5 steps and allows to detect anomalies and isolate the related original information with high accuracy. This information is very useful for the security team to reduce the time from detection to effective response. We illustrated our approach in a emulated benchmark with two semistructured sources and in a real world network security problem.

Using MBDA, identifying the timestamp and the characteristics of the anomalous traffic (\textit{e.g.}, the services affected) is a matter of seconds. We illustrate several examples in which we could identify the raw information corresponding to an attack with a level of specificity higher than  99\%, that is, more than 99\% of the isolated logs were truly related to the attack. Furthermore, extracted information corresponds in all cases to a tiny portion of far less than 1\% (a needle) of the complete data set (the haystack).

\section*{Acknowledgement}
\label{sec:Acknowledgments}
This work is partly  supported by the Spanish  Ministry of Economy and Competitiveness and FEDER funds through  project TIN2014-60346-R and TIN2017-83494-R.




\bibliographystyle{elsarticle-num}
\bibliography{3way}
\onecolumn

\include{figures}

\include{tables}

\end{document}